\documentclass[10pt,journal,twocolumns]{IEEEtran}
\usepackage{subfigure}
\usepackage{amssymb}
\usepackage{amsthm}
\usepackage{amsmath}
\usepackage{graphicx}
\usepackage{colortbl,dcolumn}
\usepackage{booktabs}
\usepackage{xcolor}
\usepackage{cite}
\usepackage{threeparttable}
\usepackage{algorithm,float}
\usepackage{algorithmic}



\newtheorem{definition}{Definition}

\newcommand{\defref}[1]{Definition~\ref{#1}}

\newcommand{\figref}[1]{Fig.~\ref{#1}}

\newcommand{\secref}[1]{Section~\ref{#1}}

\newcommand{\algref}[1]{Algorithm~\ref{#1}}

\usepackage{mathtools} 

\usepackage{enumerate}

\ifCLASSOPTIONcompsoc
\else
\fi

\ifCLASSINFOpdf
\else
\fi
\graphicspath{{pdfs/},{images/},{authorphotos/}}

\renewcommand{\hat}{\widehat}
\renewcommand{\bar}[1]{\overline{#1}}
\renewcommand{\[}{[\![ }
\renewcommand{\]}{]\!] }

\begin{document}
%
\title{ Set-Membership Filtering-Based Cooperative State
Estimation for Multi-Agent Systems }
%
%
%
%

\author{Yu~Ding,~Yirui~Cong,~\IEEEmembership{Member,~IEEE,}
        Xiangke~Wang,~\IEEEmembership{Senior Member,~IEEE,}


\IEEEcompsocitemizethanks{\IEEEcompsocthanksitem Y.~Ding, Y.~Cong, and X. Wang are with the College of Intelligence Science and Technology, National University of Defense Technology, China.\protect\\
}
}
\IEEEtitleabstractindextext{%

\begin{abstract}
In this article,  we focus on the cooperative state estimation problem of a multi-agent system.
Each agent is equipped with absolute and relative measurements.
%
The purpose of this research is to make each agent generate its own state estimation with only local measurement information and local communication with neighborhood agents using Set Membership Filter(SMF).
To handle this problem, we analyzed centralized SMF framework as a benchmark of distributed SMF and propose a finite-horizon method called OIT-Inspired centralized constrained zonotopic algorithm.
Moreover, we put forward a distributed Set Membership Filtering(SMFing) framework and develop a distributed constained zonotopic algorithm.
Finally, simulation verified  our theoretical results, that our proposed algorithms can  effectively estimate the state of each agent.
\end{abstract}

%

%

}

\maketitle
\IEEEdisplaynontitleabstractindextext

\IEEEpeerreviewmaketitle

\section{Introduction}\label{sec:Introduction}

%
%

Recent years, cooperative estimation over sensor networks has received considerable research attention  due to its extensive applications in many fields such as target tracking, environmental monitoring and industrial automation~\cite{2013Range}.
So far theories such as consensus~\cite{OlfatiSaber2009} and diffusion strategies~\cite{Bruno2018} have been used for developing filters over sensor networks.
From the purpose of cooperative estimation, there are mainly two kinds of problems.
The first one is to estimate a common target state by using multiple sensors, and this scene is common in cooperative detection~\cite{EngYuzhentao2022}.
The second one is each agent utilizes  absolute and relative measurements to make the  estimation of its own state~\cite{Viegas2017}.

For the first kind of cooperative estimation, there are a few  articles with set-membership theory to handle this problem,
which can be mainly divided into the following two types:
\begin{itemize}
\item  \textbf{Ellipsoidal SMF.}
In \cite{Maozhen2020}, a recursive resource-efficient filtering algorithm is proposed to realize distributed estimation in the simultaneous presence of the Round-Robin transmission protocol, the multi-rate mechanism and bounded noises, such that there exists a certain ellipsoid that includes all possible error states at each time instant.
Reference \cite{Zhanghao2013} considered system with sensor saturation, and derived a sufficient condition for distributed average set-membership filtering algorithm to contain the true state with recursive ellipsoids.
\item \textbf{Polytopic SMF.}
Reference \cite{Orihuela2018} developed a zonotopic estimation algorithm providing a design of obsever gains to minimize the estimation uncertainties, and proposed a method to reduce the amount of information at each time, which considered the trade-off between communication burden and estimation performance.
A distributed zonotopic filter is proposed for robust observation and FDI (Fault Detection and Isolation) of LTV (Linear Time Varying) cyber-physical systems in \cite{Combastel2018}.
The concept of bit-level reduction is introduced to relax the bit-rate requirements to communicate zonotopic sets between agents, and the negociation of data reconciliation strategies to maintain robustness/consistency under potential packet losses is proposed.
\end{itemize}

However, to the best of our knowledge, few studies paid attention to the second kind of cooperative estimation problem with set-membership filter.
Reference \cite{wang2018distributed} defined the parameterized distributed state bounding zonotope for each interconnected system, and proposed an optimization problem to minimize the effect of uncertainties based on the P-radius minimization criterion.
Similarly, \cite{Wangye2017} minimized the intersection zonotopes using an optimization-based method with a series of linear matrix inequalities (LMIs). An on-line method is also discussed for updating the correction matrices method.
These two articles are the first propose pioneer of the distributed set-membership estimation problem of systems with inertial interactions.
However, relative measurements between neighbourhood agents are not considered  in the above mentioned two studies.
In the literature, systematic studies of distributed SMFing framework are also neglected.
Thus in this article, we  study  the  centralized and distributed set-membership filtering frameworks in the cooperative estimation problem of multi-agent systems with absolute and relative measurements, such that each agent can obtain the estimate of its own state with high accuracy.

The contributions of this paper are summarized as follows:
\begin{itemize}
\item
To provide a benchmark for distributed set-membership filters, we firstly analyzed the centralized set-membership filtering framework, and put forward a finite horizon constrained zonotopic algorithm.
%
Compared with classical constrained zonotopic algorithm, our algorithm works productively.
\item
Most importantly, we propose a distributed set-membership filter framework, each can estimate its own state with  only local measurements and  measurements of its neighbours.
%
\item
A  decentralized SMF algorithm based on constrained zonotope is then developed with our distributed framework.
Finally, simulations examples verified our proposed algorithms.
%
\end{itemize}

This paper is organized as follows:
\secref{sec:Problem_Description} gives the system model and the problem description.
\secref{sec:Centralized SMF Framework and Algorithm} provides the framework of centralized SMF and our proposed finite-horizon method called OIT-Inspired constrained zonotopic algorithm.
In \secref{sec:Partially_Distributed_Set-membership_Filter}, we present the framework of distributed SMF, based on this, a distributed constrained zonotopic algorithm is developed.
Simulation results are provided to verify our theoretical results in \secref{sec:Simulation}.

Notation:
For a sample space $\Omega$, a measurable function $\mathbf{x}:\Omega \to \chi $ from the sample space $\Omega$ to a measurable set $\chi$ is called an uncertain variable.
A realization of $\mathbf{x}$ is defined as $\mathbf{x}(\omega)=x $.
The range of an uncertain variable is described by  its range:
$\[\mathbf{x} \]=\{ x(\omega):\omega \in \Omega \}$,
 $\mathbb{R}^n$ denotes the $n$- dimensional Euclidean space.
The conditional range of $\mathbf{x}$ given $\mathbf{y}=y$ is
$
\[ \mathbf{x}|y   \]=\{ \mathbf{x}(\omega) :\mathbf{y}(\omega) =y, \omega \in \Omega \}=\{\mathbf{x}(\omega) : \omega \in \Omega_{\mathbf{y}=y}     \}
$.
A directed graph $\mathcal{G}=(\mathcal{V},\mathcal{E})$ is used to represent system topology.
$\mathcal{E}=\{ (v_i,v_j): i,j \in \mathcal{V}   \} $ is the set of edges which models information flow.
%
%
%
In this paper, a set $\bar{\mathcal{N}}_i$ is defined as $\bar{\mathcal{N}}_i=i \cup \mathcal{N}_i $ to denote the union set of the $i$th agent and its in-neighbors, and $ q_i=|\bar{\mathcal{N}}_i|$, where $|\cdot|$ returns the cardinality of a given set.
The out-neighborhood of agent $i$ is denoted by $\mathcal{M}_i=\{l \in \mathcal{V} : (v_l,v_i)\in \mathcal{E} \} $.
$\mathbb{N}$ denotes the natural number set, while $\mathbb{N}_0 $ is the set of positive natural set.
$\oplus  $ stands for the Minkowski sum.
$1_N$ denotes a N-length column vector whose each element is $1$.
\section{System Model and Problem Description}\label{sec:Problem_Description}
Consider a  system composed by $N$ agents, where each agent is identified by a  positive integer $i \in \{1,2,\ldots,N \}$.
The  dynamic of agent $i$  is represented by a discrete-time  equation
\begin{align}\label{eqn:agent_dynamic}
x_{i,k+1}=A_i x_{i,k}+B_i w_{i,k}, \quad i=1,\ldots,N,
\end{align}
where $k \in \mathbb{N} $.
As a realization of $\mathbf{x}_{i,k}$,  $x_{i,k}\in \mathbb{R}_i^n$ denotes the state at time instant $k$ for agent $i$,
 $x_{i,k} \in \[ \mathbf{x}_{i,k} \] \subseteq \mathbb{R}_i^n $;
%
%
$w_{i,k}$ is  the process noise, which is  the realization of $\mathbf{w}_{i,k} \in \[\mathbf{w}_{i,k}  \] \subseteq \mathbb{R}^{p_i}  $.
$A_i \in  \mathbb{R}^{n_i \times n_i }  $  stand for the system matrix, while $B_i$ is the  input matrix with appropriate dimensions.

In this work, we consider two types of measurements, i.e., absolute measurements and relative measurements, which are widely considered in the literature ~\cite{2019Distributed}.


\textbf{Absolute measurement:}
An absolute measurement refers to agent $i$ takes an observation on its state directly,  with an observation equation as the following form:
\begin{align}\label{eqn:absolute_measurement}
y_{i,k}=C_i x_{i,k}+v_{i,k},
\end{align}
where $y_{i,k}\in \mathbb{R}^{m_i}$ is the measurement at time instant $k$ for agent $i$.
$C_i \in \mathbb{R}_i^{m_i \times n_i} $ is the measurement matrix, $v_{i,k}$ is unknown but bounded measurement noise.
 $v_{i,k} \in \[ \mathbf{v}_{i,k}  \] $,  with $d( \[\mathbf{v}_{i,k}\] ) \leq \bar{v}_i$, where $d(\cdot)$ returns the diameter of a set.

\textbf{Relative measurement:}
Relative measurements generate the observation  with both the state of agent $i$ the in-neighborhood agents $j \in \mathcal{N}_i$,
\begin{align}\label{eqn:relative_measurement}
z_{i,j,k}=D_i(x_{i,k}-x_{j,k})+r_{i,j,k},~j\in \mathcal{N}_i,
\end{align}
where $z_{i,j,k}\in \mathbb{R}^{p_i}$ is the measurement at time instant $k$ between agent $i$ and its in-neighborhood agent $j$.
$D_i \in \mathbb{R}_i^{p_i \times n_i}  $ is the  measurement matrix.
$r_{i,j,k} \in \[  \mathbf{r}_{i,j,k}  \]$ is the measurement noise, with $d(\[ \mathbf{r}_{i,j,k}\]) \leq \bar{r}_{i,j}$.

%
In this article, we consider the communication topology is consistent with the time-invariant measurement topology, i.e., $(v_i,v_j)\in \mathcal{E} $ implies agent $i$ takes a relative measurement from agent $j$ and receives information from agent $j$ simultaneously.
%

The aim of this paper is to derive an estimate of uncertain range $S_{i,k} \supseteq \[\mathbf{x}_{i,k}|y_{0:k},z_{0:k}  \]$ for each agent $i$  by using all the  measurements $y_{i,k} $, $y_{j,k}$ and $z_{i,j,k},~j\in \mathcal{N}_i$, such that at each instant $k$, $x_{i,k} \in S_{i,k} $.
%

%


\section{Centralized SMFing Framework and Algorithm}\label{sec:Centralized SMF Framework and Algorithm}

In this section, a centralized SMFing framework is  presented as the benchmark.
Based on this framework, we propose a finite-horizon SMFing constrained zonotopic algorithm with high efficiency.
%
%
Let $x_{k}=[x_{1,k}^T,x_{2,k}^T,\cdots,x_{N,k}^T ]^T$
and
$w_{k}=[w_{1,k}^T,w_{2,k}^T,\cdots,w_{N,k}^T ] ^T$.
The system dynamics can be written as follows:
\begin{align}
x_{k+1}&=A_c x_{k}+B_c w_{k},
\end{align}
where
\begin{align}
 A_c&=\mathrm{diag}\{A_1,~A_2,~\cdots,~A_N\}, \\ \notag
  B_c&=\mathrm{diag}\{B_1, ~B_2, \cdots, ~B_N\}.
 \end{align}

%
The augmented form of  absolute measurements is
\begin{align}
y_{k}&=[y_{1,k}^T,y_{2,k}^T,\cdots,y_{N,k}^T ]^T  \\ \notag
              &=C_c x_k+ v_{k},
\end{align}
where
\begin{align}
C_c&=\mathrm{diag}\{C_1,~C_2,\cdots,~C_N   \}, \\ \notag
v_{k}&=[v_{1,k}^T, v_{2,k}^T,\cdots,v_{N,k}^T]^T.
\end{align}
%
%

Since the relative measurements of an agent are related to its in-neighbors, we define $z_{i,k}$ as the all the relative measurements taken by agent $i$, i.e.,
\begin{align}
z_{i,k}&=D_c \begin{bmatrix}  x_{i,k}-x_{j_1,k}  \\
                      x_{i,k}-x_{j_2,k}   \\
                      \vdots \\
                      x_{i,k}-x_{j_s,k}  \end{bmatrix}
  +\begin{bmatrix} r_{i,j_1,k} \\   r_{i,j_2,k} \\ \vdots \\r_{i,j_s,k}    \end{bmatrix}
     , \quad j_1,\cdots,j_s\in \mathcal{N}_i,
\end{align}
where $D_c=\mathrm{diag}\{D_1,~D_2,\cdots,~D_N  \} $.

The relative measurements for all the agents  can be written as a compact form
$z_{k}=[z_{1,k}^T,z_{2,k}^T,\cdots,z_{N,k}^T]^T$.

Let $Y_{k}=\begin{bmatrix} y_{k} \\ z_{k}    \end{bmatrix} $, $r_{i,k}=[ r_{i,j_1,k}^T, ~ r_{i,j_2,k}^T,\cdots,~ r^T_{i,j_s,k} ]^T$,
$V_{k}=[ v_k^T , r_{1,k}^T , r_{2,k}^T , \cdots , r_{N,k}^T ]^T$.
The augment form of all the measurements can be written as
\begin{align}\label{eqn:centralized_augmented_observation}
Y_k&=\begin{bmatrix}C_c \\ D_c \end{bmatrix} x_k+\begin{bmatrix}   v_k  \\ r_{1,k}  \\ \vdots \\ r_{N,k}\end{bmatrix} \\ \notag
            &=H_c x_k +V_k.
\end{align}

The prediction and update of classical centralized set-membership filter are given as follows:
 \begin{align}
    &\[\mathbf{x}_k|Y_{0:k-1} \] =A_c \[\mathbf{x}_{k-1}|Y_{0:k-1} \] \oplus  B_c \[\mathbf{w}_{k-1}  \], \label{eqn:centralized_prediction}
     \\
   & \[\mathbf{x}_k |Y_{0:k} \]=[\underset{V_k \in \[\mathbf{V}_k\]}{\bigcup} H^{-1}_{c}(Y_k -V_k) ] \cap \[\mathbf{x}_k| Y_{0:k-1} \], \label{eqn:centralized_update}
    \end{align}
where $H^{-1}_{c}(\cdot)$ is the inverse map of $H_c(\cdot)$.
The uncertain  set of agent $i$  solved by centralized set-membership filter is denoted by $ S_{i,k}^c=\{x_{i,k} :(x_{1,k}, x_{2,k},\ldots, x_{N,k} ) \in \[\mathbf{x}_k|Y_{0:k}\]  \}$.

A common  geometric figure to realize this centralized framework is the extended constrained zonotope, which is defined as follows.
\begin{definition}  (Extended Constrained Zonotope~\cite{2022YiruiCong}) \label{def:Extended_Constrained_Zonotope}
A set $\mathcal{Z} \subseteq \mathbb{R}^n$ is an extended constrained zonotope if there exists a quintuple $(G,c,A,b,h) \in \mathbb{R}^{ n \times n_g} \times \mathbb{R}^n \times \mathbb{R}^{ n_c \times n_g} \times \mathbb{R}^{ n_c} \times [0,\infty ]^{n_g} $ such that $\mathcal{Z}$ is expressed by
\begin{align}\label{eqn:Extended_Constrained_Zonotope}
\{ G \xi +c : A \xi =b, \xi \in \underset{j=1}{\overset{n_g}{\prod}} [-h^{(j)}, h^{(j)} ] \} =:Z(G,c,A,b,h),
\end{align}
where $ h^{(j)}$ is the $j$\textsuperscript{th} component of $h$.
\end{definition}
When $h^{(j)}=1,~j=1,\cdots,n_g$, \defref{def:Extended_Constrained_Zonotope} becomes the classical constrained zonotope in \cite{2016Constrained}.
$\hat{Z}_k^c=(G^c,c^c,A^c,b^c,h^c) \supseteq  \[x_k | Y_{0:k}  \]$ denotes the centralized posterior uncertain range of the augmented system state at  $k$.
%

Since the centralized method is optimal, we can employ it as a benchmark.
However, the complexity  will go unbounded as time elapses.
To handle this problem, we propose a finite-horizon algorithm (see \algref{alg:OIT Inspired Centralized Constrained Zonotope Set-membership Algorithm})  based on the theoretical result of ~\cite{2022YiruiCong}, and we provide a line by line explanation as follows:
%

\begin{algorithm}
\caption{OIT Inspired Centralized Constrained Zonotopic SMF Algorithm}
\label{alg:OIT Inspired Centralized Constrained Zonotope Set-membership Algorithm}
\begin{algorithmic}[1]
\STATE $\mathbf{Initialization} $. Bounded Constrained zonotope $\[ \mathbf{x}_0\] \subset \mathbb{R}^{n \times N} $, $\bar{\delta} \geq \mu_0-1$.
\STATE \textbf{if} $k \leq \bar{\delta} $ \textbf{then}
\STATE $\hat{Z}_k^c \leftarrow$
 ~\eqref{eqn:centralized_prediction} and ~\eqref{eqn:centralized_update}.
\STATE \textbf{else}
\STATE Set $\hat{Z}_{k-\bar{\delta}}^{c}=\mathbb{R}^{n \times N}$, with ~\eqref{eqn:centralized_prediction} and ~\eqref{eqn:centralized_update}, solve $\hat{Z}_k^c$.
\end{algorithmic}
\end{algorithm}

\begin{itemize}
\item Line 1 initializes ~\algref{alg:OIT Inspired Centralized Constrained Zonotope Set-membership Algorithm},
where the additional parameter $\bar{\delta}$ is the length of time slide window, that a larger $\bar{\delta}$ leads to higher accuracy but increases the complexity.
$\mu_0$ is the observability index of the augmented system.
\item Line 3 is for $k \leq \bar{\delta}$, which is the same as the classical constrained zonotope SMF algorithm~\cite{2021Rethinking}.
\item Line 5  is for $k>\bar{\delta} $, which is a finite-horizon version of   Section 5 of ~\cite{2016Constrained} over the time window $[k-\bar{\delta},k ] $;
it is a simplified version of the OIT inspired constrained zonotopic SMF in~\cite{2022YiruiCong}.
\end{itemize}

%
%

\section{ Distributed SMFing Framework and Constrained Zonotopic Algorithm}\label{sec:Partially_Distributed_Set-membership_Filter}
In this section, we propose a framework of distributed set-membership filtering framework that each agent only utilizes its local observation and communication to realized its own high-accuracy state estimation.
Based on this framework, we develop a distributed constrained zonotopic  SMF algorithm.
\subsection{Distributed SMFing Framework}
Without loss of generality, we assume agents $j_1,~j_2,~\cdots,~j_s$ are the in-neighborhood agents of  $i$.
Let $x_{\bar{\mathcal{N}}_i,k}=[x^T_{i,k}, x^T_{j_1,k},\cdots,x_{j_s,k}^T ]^T$ and  $W_{\bar{\mathcal{N}}_i,k}=[ w_{i,k} ^T, w_{j_1,k} ^T, \cdots , w_{j_s,k}^T ]^T$ with $W_{\bar{\mathcal{N}}_i,k} \in \[\mathbf{W}_{\bar{\mathcal{N}}_i,k}\] $.
Then the system dynamic and measurement equation can be written as
\begin{align}
x_{\bar{\mathcal{N}}_i,k+1}&=\begin{bmatrix}A_i  & & &  \\
                  & A_{j_1} & & \\
                  & & \ddots & \\
                  & & &   A_{j_s} \end{bmatrix}
                  \begin{bmatrix}
                 x_{i,k}  \\  x_{j_1,k}  \\ \vdots \\ x_{j_s,k} \end{bmatrix}
                 \\  \notag
                  &+\begin{bmatrix}B_i  & & &  \\
                  & B_{j_1} & & \\
                  & & \ddots & \\
                  & & &   B_{j_s} \end{bmatrix}
         \begin{bmatrix} w_{i,k} \\  w_{j_1,k} \\ \vdots \\ w_{j_s,k}  \end{bmatrix} \\ \notag
       &= A_{\bar{\mathcal{N}}_i} x_{\bar{\mathcal{N}}_i,k}  +B_{\bar{\mathcal{N}}_i} W_{\bar{\mathcal{N}}_i,k},
\end{align}

\begin{align}
Y_{\bar{\mathcal{N}}_i,k}&=\begin{bmatrix} C_{i}  x_{i,k} \\ C_{j_1} x_{j_1,k}  \\ \vdots \\ C_{j_s,k} x_{j_s,k}  \\ D_i (x_{i,k}-x_{j_1,k} ) \\ \vdots \\ D_i(x_{i,k}-x_{j_s,k} ) \end{bmatrix}
+\begin{bmatrix} v_{i,k} \\ v_{j_1,k} \\ \vdots \\ v_{j_s,k} \\ r_{i,j_1,k} \\ \vdots \\ r_{i,j_s,k} \end{bmatrix} \\ \notag
&= H_{\bar{\mathcal{N}}_i,k} x_{\bar{\mathcal{N}}_i,k} +V_{\bar{\mathcal{N}}_i,k},
\end{align}
where $V_{\bar{\mathcal{N}}_i,k}=[v_{i,k}^T,v_{j_1,k}^T,\cdots,v_{j_s,k}^T,r_{i,j_1,k}^T,\cdots,r_{i,j_s,k}^T ] ^T $.

For each agent $i$,  $\bar{S}_{i,k}^d$ and $S_{i,k}^d$ denote the prior and posterior uncertain set at $k$, respectively.
The joint uncertain prior and posterior range of $(x_{i}, x_{j_1},\cdots,x_{j_s} ) $ at  $k$ are denoted by $ \bar{S}_{i,j_1,j_2,\cdots,j_s,k}^d$ and $ S_{i,j_1,j_2,\cdots,j_s,k}^d$.
For convenience, we simplify them as $\bar{S}_{ \bar{\mathcal{N}}_i,k}^d $ and $S_{\bar{\mathcal{N}}_i,k}^d $, respectively.
The framework of joint distributed set-membership filter is proposed in Algorithm 2 and its detailed explanation is given as follows:

\begin{algorithm}\label{alg:Partially_Distributed_Set-membership_Filter_Framework1}
\caption{Distributed Set-membership Filtering Framework}
\begin{algorithmic}[1]
\STATE {$\mathbf{Initialization} $. For agent $i$, set the initial joint prior range $S_{i,0}^d$
}
\LOOP
\STATE $\mathbf{Prediction}$. Given $S_{i,k-1}^d$, each agent's prior range $\bar{S}_{i,k}^d$ is given by
       \begin{align}
       \bar{S}_{i,k}^d=A_i(S_{i,k-1}^d) \oplus B_i \[ \mathbf{w}_{i,k-1}  \] , \quad i=1,2,\cdots,N
       \end{align}
\STATE Receive $\bar{S}_{j,k}^d$ from neighborhood agents
       The joint prior uncertain range $\bar{S}_{\bar{\mathcal{N}}_i,k}^p=\bar{S}_{i,k}^d \times \underset{j \in \mathcal{N}_i}{\prod} \bar{S}_{j,k}^d $.
\STATE $\mathbf{Measurement}$. Take integrated measurements $Y_{\bar{\mathcal{N}}_i,k}$ .
\STATE $\mathbf{Joint~ Update}$. For $k \in \mathbb{N}_0$, given the observation $Y_{\bar{\mathcal{N}}_i,k}$ the joint posterior set  $S_{\bar{\mathcal{N}}_i,k}^{d}$ is formulated by
    \begin{align}
    S_{\bar{\mathcal{N}}_i,k}^{d}=[\underset{V_{i,k} \in \[\mathbf{V}_{i,k} \] }{\bigcup} H_{\bar{\mathcal{N}}_i,k}^{-1}(Y_{\bar{\mathcal{N}}_i,k}-V_{\bar{\mathcal{N}}_i,k})     ]  \cap
    \bar{S}_{\bar{\mathcal{N}}_i,k}^d
    \end{align}
\STATE Receive $S_{l,k}^{d,i}=\{x_{i,k}: (x_{l,k}, x_{i,k},\cdots, x_{l_s,k}   ) \in S_{\bar{\mathcal{N}}_l,k}^{d}   \}  $ from $l \in \mathcal{M}_i \cap \mathcal{N}_i $, $l_1,\cdots,l_s \in \mathcal{N}_l$.
If $\mathcal{M}_i \cap \mathcal{N}_i=\emptyset$, $S_{l,k}^{d,i}=R^n_i$.
\STATE   $ \mathbf{Update~ Intersection}$.
The joint posterior uncertain range of UAV $i$ is formulated by
\begin{align}
S_{i,k}^d= [ \underset{l \in \mathcal{M}_i \cap \mathcal{N}_i}{\cap} S_{l,k}^{d,i}] \cap [\underset{\mathbb{R}_i^n \times \underset{j\in \mathcal{N}_i}{\prod} \mathbb{R}_{j}^n \to \mathbb{R}_i^n  }{\mathrm{Proj}}  S_{\bar{\mathcal{N}}_i,k}^{d}  ]
\end{align}
\STATE $k=k+1$;
\ENDLOOP
\end{algorithmic}
\end{algorithm}

\begin{itemize}
\item Line 1 initializes the $i$\textsuperscript{th} agent and its in-neighborhood agents' joint uncertain range $S_{\bar{\mathcal{N}}_i,0}^d$.
\item Line 4 denotes the communication process that transmits the information required in the joint progress.
%
\item Line 6 calculates the joint prediction $\bar{S}_{\bar{\mathcal{N}}_i,k}^d$.
\item Line 7-8 formulate   $S_{\bar{\mathcal{N}}_i,k}^{d*} $ using a projection method, where the state estimation can be implemented by computing the intersection of  the  posterior uncertain range of agents that receive information from  $i$.
\end{itemize}

\subsection{Distributed  Constrained Zonotopic SMF Algorithm}\label{sec:Distributed SMF Zonotopic Algorithm}
In this section, we design the distributed SMF algorithm based on constrained zonotope method.
To distinguish with the system dynamic matrix $A_i$, in this section we use $\bar{A}_{i,k}$ and  $\hat{A}_{i,k}$  to denote prior and posterior in the constrained zonotope expression, respectively.
The calculation in each step is highlighted as follows:

(1) Initialization.
For agent $i$, set the initial uncertain range as
\begin{align}\label{eqn:distributed_CZ_initialization}
 &\hat{Z}_{i,0}^d= (\hat{G}_{i,0}^d,\hat{c}_{i,0}^d,\hat{A}_{i,0}^d,\hat{b}_{i,0}^d,\hat{h}_{i,0}^d) \\ \notag
 &= \{z:z=\hat{G}_{i,0}^d \xi +\hat{c}_{i,0}^d, \hat{A}_{i,0}^d\xi =\hat{b}_{i,0}^d,
 \xi \in   \underset{t=1}{\overset{n_g^i}{\prod} }  [-\hat{h}_{i,0}^{d,t}, \hat{h}_{i,0}^{d,t}] \},
 \end{align}
where $\hat{h}_{i,0}^{d,t}$ denotes the $t$\textsuperscript{th} component of $\hat{h}_{i,0}^d$.
%


(2) Prediction.
Given the  uncertain range of last instant $\hat{Z}_{i,k-1}^d=(\hat{G}_{i,k-1}^d,\hat{c}_{i,k-1}^d,\hat{A}_{i,k-1}^d,\hat{b}_{i,k-1}^d,\hat{h}_{i,k-1}^d) $,
 the prior uncertain range at $k$ is given by
 \begin{align}\label{eqn:local_prediction}
 &\bar{Z}_{i,k}^d=(\bar{G}_{i,k}^d,\bar{c}_{i,k}^d,\bar{A}_{i,k}^d,\bar{b}_{i,k}^d,\bar{h}_{i,k}^d ) \\ \notag
 &=([A_{\bar{\mathcal{N}}_i} \bar{G}_{i,k}^d \quad B_i \bar{G}_{w_{i,k}}], A_{\bar{\mathcal{N}}_i} \bar{c}_{i,k}^d ,
 \begin{bmatrix}\bar{A}_{i,k}^d & 0 \\ 0 & \bar{A}_{w_{i,k}}\end{bmatrix}, \\ \notag
& \begin{bmatrix}\bar{b}_{i,k}^d \\ \bar{b}_{w_{i,k}} \end{bmatrix},
 \begin{bmatrix}\bar{h}_{i,k}^d \\ \bar{h}_{w_{i,k}} \end{bmatrix} ).
 \end{align}
The joint prior uncertain range is given by
\begin{align}\label{eqn:prior_cartesian_product}
&\bar{Z}_{\bar{\mathcal{N}}_i,k}^d=(\bar{G}_{\bar{\mathcal{N}}_i,k}^d,\bar{c}_{\bar{\mathcal{N}}_i,k}^d,\bar{A}_{\bar{\mathcal{N}}_i,k}^d,\bar{b}_{\bar{\mathcal{N}}_i,k}^d,\bar{h}_{\bar{\mathcal{N}}_i,k}^d) \\ \notag
&=(\mathrm{diag}\{ \bar{G}_{i,k}^d,~\bar{G}_{j_1,k}^d,\cdots,~\bar{G}_{j_s,k}^d \}, \begin{bmatrix}\bar{c}_{i,k}^d \\ \vdots \\ \bar{c}_{j_s,k}^d \end{bmatrix},  \\ \notag
&\mathrm{diag}\{ \bar{A}_{i,k}^d,~\bar{A}_{j_1,k}^d,\cdots,~\bar{A}_{j_s,k}^d \},\begin{bmatrix}\bar{b}_{i,k}^d \\ \vdots \\ \bar{b}_{j_s,k}^d \end{bmatrix},\begin{bmatrix}\bar{h}_{i,k}^d \\ \vdots \\ \bar{h}_{j_s,k}^d \end{bmatrix})
\end{align}

(3) Joint update.
Given Sample integrated measurements at instant $k$ $y_{i,k}$ and $z_{i,k}$, joint update is given by
\begin{align}\label{eqn:distributed_CZ_update}
&\hat{Z}_{\bar{\mathcal{N}}_i,k}^{d}=(\hat{G}_{\bar{\mathcal{N}}_i,k}^d, \hat{c}_{\bar{\mathcal{N}}_i,k}^d,\hat{A}_{\bar{\mathcal{N}}_i,k}^d,\hat{b}_{\bar{\mathcal{N}}_i,k}^d,
\hat{h}_{\bar{\mathcal{N}}_i,k}^d ) \\ \notag
&=([\bar{G}_{\bar{\mathcal{N}}_i,k}^p ~ 0 ],\bar{c}_{\bar{\mathcal{N}}_i,k}^d,
    \begin{bmatrix}\bar{A}_{\bar{\mathcal{N}}_i,k}^d & 0 \\ 0 &  \bar{A}_{\bar{\mathcal{N}}_i,k}^{V_i} \\ \bar{G}_{\bar{\mathcal{N}}_i,k}^d & (H_{\bar{\mathcal{N}}_i,k}^T H_{\bar{\mathcal{N}}_i,k} )^{-1} H_{\bar{\mathcal{N}}_i,k} G_{\bar{\mathcal{N}}_i,k}^{V_i} \end{bmatrix}  \\ \notag
&   , \begin{bmatrix} \bar{b}_{\bar{\mathcal{N}}_i,k}^d \\ b_{\bar{\mathcal{N}}_i,k}^{V_i} \\ \bar{c}_{\bar{\mathcal{N}}_i,k}^d+(H_{\bar{\mathcal{N}}_i,k}^T H_{\bar{\mathcal{N}}_i,k} )^{-1} H_{\bar{\mathcal{N}}_i,k}Y_{i,k} \end{bmatrix} , \begin{bmatrix} h_{\bar{\mathcal{N}}_i,k}^d \\ h_{\bar{\mathcal{N}}_i,k}^{V_i} \end{bmatrix}).
\end{align}

(4) Update intersection.
Let $e_{\alpha,u}$ denote a projection matrix to choose the $\alpha $th element from a $u$-length vector, and $E_{\alpha,u}=e_{\alpha,u} \otimes I_n$.
%

The final  uncertain range at instant $k$ after update intersection $\hat{Z}_{i,k}^d=(\hat{G}_{i,k}^d, \hat{c}_{i,k}^d, \hat{A}_{i,k}^d, \hat{b}_{i,k}^d,\hat{h}_{i,k}^d )$ is formulated by
\begin{align}
&\hat{G}_{i,k}^d=[E_{1,q_i} \hat{G}_{\bar{\mathcal{N}}_i,k} \quad 0 ],   \label{eqn:distributed_CZ_update_intersection_G} \\
&\hat{c}_{i,k}^d= E_{1,q_i} \hat{c}_{\bar{\mathcal{N}}_i,k},    \label{eqn:distributed_CZ_update_intersection_c}  \\
&\hat{A}_{i,k}^d=\begin{bmatrix} A_{i,k}^{1,1} \\ A_{i,k}^{2,1} \end{bmatrix},\label{eqn:distributed_CZ_update_intersection_A}
    \\
&   \hat{b}_{i,k}^d=\begin{bmatrix}B_{i,k}^{1,1} \\  B_{i,k}^{2,1} \end{bmatrix},
\\
&\hat{h}_{i,k}^d=  \begin{bmatrix}\hat{h}^d_{\bar{\mathcal{N}}_i,k} \\ \hat{h}^d_{\bar{\mathcal{N}}_{l_1},k} \\ \vdots
  \\\hat{h}^d_{\bar{\mathcal{N}}_{l_s},k} \end{bmatrix},    \label{eqn:distributed_CZ_update_intersection_h}
   \end{align}
 where $ A_{i,k}^{1,1}=\mathrm{diag}_{l \in M_i \cap \bar{\mathcal{N}}_i}\{\hat{A}_{\bar{\mathcal{N}}_l,k}^d \} $ ,
 and $A_{i,k}^{2,1}=[A^{1,1} \quad  A^{1,2} ]$ is a  block matrix composed by $A^{1,1}=\mathbf{1}_{q_i } \otimes  E_{1,q_i} \hat{G}_{\bar{\mathcal{N}}_i,k}^d $ with  $q_i=| \bar{\mathcal{N}}_i  |$;
  $E_{\alpha_l,q_l}$ denotes the serial number of $i$ in  $\bar{\mathcal{N}}_l$;
 $A^{1,2}=\mathrm{diag}_{l \in M_i \cap \bar{\mathcal{N}}_i} \{E_{\alpha_l, q_l} \hat{G}_{\bar{\mathcal{N}}_{l},k}^d \} $,
$B_{i,k}^{1,1}=\begin{bmatrix}  \hat{b}_{\bar{\mathcal{N}}_i,k}\\ \vdots \\ \hat{b}_{\bar{\mathcal{N}}_{j_s},k} \end{bmatrix}$,
and $B_{i,k}^{2,1}=\begin{bmatrix}E_1  c_{i,j \in \mathcal{N}_i,k} -E_{\alpha_1}  c_{l_1,m \in \mathcal{N}_{l_1},k}
  \\ \vdots
  \\ E_1  c_{i,j \in \mathcal{N}_i,k} -E_{\alpha_{l_s}}  c_{l_s,m \in \mathcal{N}_{l_s},k}
  \end{bmatrix}$.

\begin{algorithm}
\caption{Distributed Constrained Zonotopic SMF Algorithm}
\label{alg:Partially_Distributed_Set-membership_Filter_Algorithm_Based_on_Constrained_Zonotope}
\begin{algorithmic}[1]
\STATE  The initial uncertain range of agent $i$ is given by ~\eqref{eqn:distributed_CZ_initialization}.
\LOOP
\STATE  The joint prior uncertain range of agent $i$ and its in-neighborhood agents at instant $k$ is given by ~\eqref{eqn:prior_cartesian_product}.
\STATE  The joint posterior uncertain range is updated as ~\eqref{eqn:distributed_CZ_update}.
\STATE The  posterior uncertain range of agent $i$ at instant $k$ is formulated by $\hat{Z}_{i,k}^{d*}=(\hat{G}_{i,k}^{d*},\hat{c}_{i,k}^{d*},\hat{A}_{i,k}^{d},\hat{b}_{i,k}^{d},\hat{h}_{i,k}^{d})$, where each matrix is given in \eqref{eqn:distributed_CZ_update_intersection_G} - \eqref{eqn:distributed_CZ_update_intersection_h}.
\STATE Find an interval hull $(\hat{G}_{i,k}^d,\hat{c}_{i,k}^d)$ such that $\hat{Z}_{i,k}^d=\{ z=\hat{G}_{i,k}^d \xi +\hat{c}_{i,k}^d \} \supseteq Z_{i,k}^{d*}  $,
  and the final posterior uncertain range at instant $k$ is $\hat{Z}_{i,k}^d=(\hat{G}_{i,k}^d,\hat{c}_{i,k}^d,[~],[~],B_{\infty}) $.
\STATE $k=k+1$;
\ENDLOOP
\end{algorithmic}
\end{algorithm}

Taking the topology in ~\figref{fig:system_topology} as an example, we explain the update intersection step in Line 6 of ~\algref{alg:Partially_Distributed_Set-membership_Filter_Algorithm_Based_on_Constrained_Zonotope} as follows:
\begin{itemize}
\item     The augmented  state of UAV2 and its in-neighbourhood UAVs can be written as $\mathbf{x}_{\bar{\mathcal{N}}_2,k }=[\mathbf{x}_{2,k}^T,\mathbf{x}_{1,k}^T,\mathbf{x}_{3,k}^T,\mathbf{x}_{4,k}^T ]^T$, $q_2=|\bar{\mathcal{N}}_2 |=4 $.
    Clearly, $\mathbf{x}_{2,k}=[I \quad 0 \quad 0 \quad 0 ]\mathbf{x}_{\bar{\mathcal{N}}_2,k } $, the projection matrix is $E_{1,4}=[I \quad 0 \quad 0 \quad 0 ]$.
\item  Notice that UAV2 is also the in-neighbourhood agent of UAV4, which implies $\mathbf{x}_{2,k}$ will appear in the augmented vector $\mathbf{x}_{\bar{\mathcal{N}}_4,k }$, i.e., $\mathbf{x}_{2,k}=[0 \quad I  \quad 0 ] [\mathbf{x}_{4,k}^T, \mathbf{x}^T_{2,k},\mathbf{x}^T_{3,k} ]^T $, $q_4=|\bar{\mathcal{N}}_4 |=3$.
    The corresponding projection matrix is $E_{2,3}=[0 \quad I \quad 0 ]$.
    The posterior range after update intersection $\hat{Z}_{2,k}^d $ is given by
    \begin{align}
    &\hat{G}_{2,k}^d=[E_{1,4}\hat{G}^d_{\bar{\mathcal{N}}_2,k} \quad 0 ], \\ \notag
    &\hat{c}_{2,k}^d=E_{1,4}\hat{c}^d_{\bar{\mathcal{N}}_2,k}, \\ \notag
    &\hat{A}_{2,k}^d=\begin{bmatrix} \hat{A}^d_{\bar{\mathcal{N}}_2,k} & 0 \\ 0 & \hat{A}^d_{\bar{\mathcal{N}}_4,k}
     \\ E_{1,4} \hat{G}^d_{\bar{\mathcal{N}}_2,k} & -E_{2,3} \hat{G}^d_{\bar{\mathcal{N}}_4,k} \end{bmatrix}, \\  \notag
    & \hat{b}_{2,k}^d= \begin{bmatrix} \hat{b}^d_{\bar{\mathcal{N}}_2,k} \\ \hat{b}^d_{\bar{\mathcal{N}}_4,k} \\
    E_{1,4} \hat{c}^d_{\bar{\mathcal{N}}_2,k}  -E_{2,3} \hat{c}^d_{\bar{\mathcal{N}}_4,k}\end{bmatrix}, \\   \notag
    &\hat{h}_{2,k}^d=\begin{bmatrix}\hat{h}^d_{\bar{\mathcal{N}}_2,k} \\ \hat{h}^d_{\bar{\mathcal{N}}_4,k} \end{bmatrix}.
    \end{align}
\end{itemize}

In Line 8,  we take an interval hull instead of the precise posterior constrained zonotope for fast calculation.

\section{Simulation Results}\label{sec:Simulation}
In this section, we consider  five UAVs in a 2-D plane, each utilizes  absolute and relative measurements to localize its position.
Let $x_{i,k}=[p^T_{x,i,k},v^T_{x,i,k},p^T_{y,i,k},v^T_{y,i,k} ]^T$, where $p^T_{i,k} $ and $v^T_{i,k} $ denote the position and velocity of UAV $i$ at instant $k$, respectively.
The parameters are listed as follows:
\begin{align}
A_{i,k}=\begin{bmatrix}a_{11} & a_{12} \\ a_{21} & a_{22} \end{bmatrix}
\end{align}
with $a_{11}= a_{22}=1+\frac{1}{\omega}(\mathrm{sin}(k+1)\omega T-\mathrm{sin} (k\omega T) ) $, \\ $a_{12}=-a_{21}=-\frac{1}{\omega}(\mathrm{cos}(k+1)\omega T-\mathrm{cos} (k\omega T) )$, and
\begin{align}
B_i=\begin{bmatrix}\frac{T^2}{2} \\ T \end{bmatrix},
C_i=[1 \quad 0],
D_i=I_2,\omega=1, T=\frac{\pi}{12}
\end{align}

The topology  is shown in ~\figref{fig:system_topology} and the noise constrained zonotope is  as follows:
\begin{figure}
  \centering
  \includegraphics[width=0.8\columnwidth]{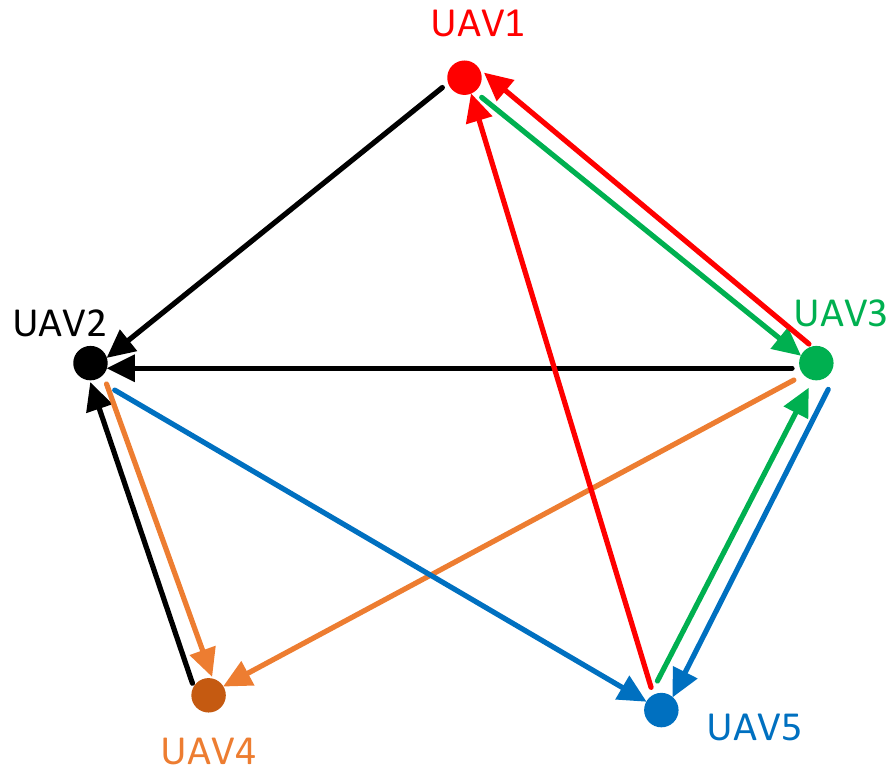}
  \caption{ System measurement and communication topology. }\label{fig:system_topology}
\end{figure}

\begin{align}
\[ w_{i,k}\]=\[r_{i,j,k}\]  =[-1,1] \times [-1,1], \[ v_{i,k} \]=[-1,1].
\end{align}

The initial state and its uncertain range are randomly generated by Matlab.

\begin{figure}[h]
\centering
\subfigure[]{\includegraphics [width=0.9\columnwidth, trim = 0 0 0 0]{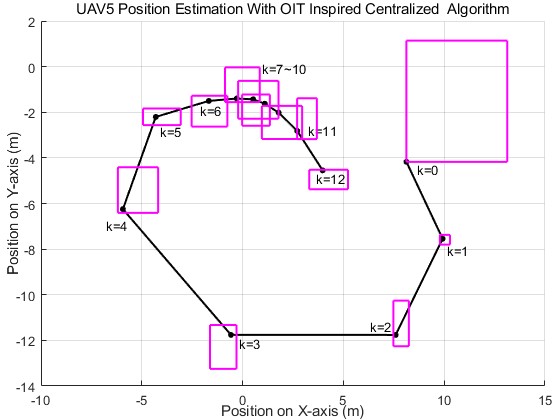}\label{fig:OIT_Centralized_SMF_UAV5}}\\
\subfigure[]{\includegraphics [width=0.9\columnwidth, trim = 0 0 0 0]{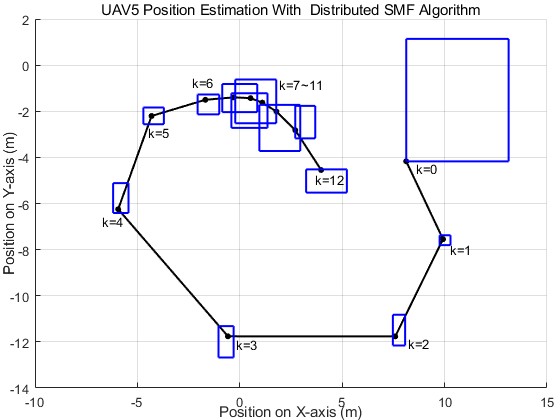}\label{fig:Distibuted_SMF_Algorithm_UAV5}}
\caption{Position estimation of UAV 5 with OIT-Inspired centralized and distributed SMF algorithm.
(a)UAV 5 position estimation sketch map with OIT-Inspired centralized constrained zonotopic algorithm.
(b) UAV5 position estimation with proposed distributed SMF constrained zonotopic algorithm.
}\label{fig:UAV5_algorithm_verify}
\end{figure}

Due to the page limit, we only present the results of UAV 4 and UAV 5.
In ~\figref{fig:UAV5_algorithm_verify}, the trajectory of  UAV5 is the black line, with the dots representing the real positions at different time steps.
The position estimation set in 2-D plane are rectangles.
The constrained zonotope generated by OIT-Inspired centralized SMF zonotopic algorithm is presented by pink rectangles, shown as ~\figref{fig:OIT_Centralized_SMF_UAV5}.
The interval hull corresponding to the distributed SMF algorithm is plotted by blue rectangles; see ~\figref{fig:Distibuted_SMF_Algorithm_UAV5}.
At each step, We can see that the constrained zonotope presented by both the OIT centralized  algorithm and distributed SMF algorithm  contain the real positions of UAV 5, which corroborates the effectiveness of our proposed methods.

\begin{figure}[ht]
\centering
\subfigure[]{\includegraphics [width=0.9\columnwidth, trim = 0 0 0 0]{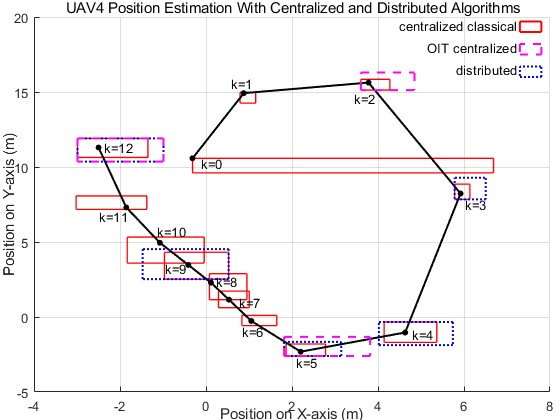}\label{fig:Distibuted_Set-membership_UAV4}}\\
\subfigure[]{\includegraphics [width=0.9\columnwidth, trim = 0 0 0 0]{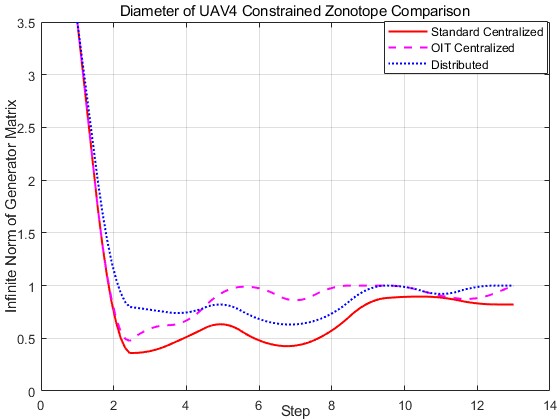}\label{fig:Gnorm_comparison_UAV4}}
\caption{The comparison of standard centralized, OIT-Inspired centralized algorithm and distributed SMF constrained zonotopic algorithm.
(a) The constrained zonotope generated by standard centralized SMF algorithm,
OIT Inspired centralized algorithm and distributed SMF algorithm simultaneously.
(b) The infinite norm of generator matrix $\| G \|_{\infty}$ corresponding to the diameter of uncertain range.
}\label{fig:UAV4_algorithm_comparison}
\end{figure}

In ~\figref{fig:Distibuted_Set-membership_UAV4}, we compare the proposed OIT Inspired centralized algorithm and distributed SMF algorithm with the benchmark (the standard centralized algorithm).
%
The constrained zonotopes generated by standard centralized SMF algorithm  each step  are presented with red solid line rectangles.
The OIT Inspired centralized  zonotopes are depicted by pink dash line rectangles.
Zonotopic rectangles presented by our proposed distributed algorithm is plotted  by blue dotted line.
At instant $k=5$ and $k=12$, the three rectangles are labeled together.
The red rectangle, corresponding to the standard centralized algorithm, is the smallest among three boxes, which implies both OIT-Inspired method and distributed method have conservativeness.
To measure the conservativeness directly, we consider using the diameter of  sets, i.e.,
%
\begin{align}\label{eqn:diameter_of_set}
d(S)=\underset{s_1,s_2 \in S  }{\mathrm{sup}}\|s_1-s_2  \|_{\infty}.
\end{align}

For an interval hull in 2-D plane, $d(S)$ is the maximum edge length, which is the twice of the  infinite norm of  $\hat{G}$ matrix.
Thus in ~\figref{fig:Gnorm_comparison_UAV4} we execute comparison of  infinite norm of $\hat{G}$ matrix of the interval hull computed by three algorithms.
From ~\figref{fig:Gnorm_comparison_UAV4} we can see that the curve corresponding to OIT-Inspired centralized method and distributed method is always above the standard centralized method;
This implies our OIT inspired centralized algorithm sacrifice over-estimation accuracy to obtain low complexity, and our distributed SMF method realize distributed structure with low conservativeness.
%

%
%

\section{Conclusion}\label{sec:Conclusion}
In this paper, we study the state estimation problem of a multi-agent system with absolute and relative measurements.
Firstly, we analyzed the centralized SMF framework as the benchmark.
%
To restrict the unlimited complexity increasing in the classical constrained zonotopic algorithm, we develop a finite-horizon version called OIT-Inspired centralized algorithm
Secondly, a distributed SMFing framework is presented. Utilizing this framework, each agent can estimate its own state with local measurements and communications with neighbourhood agents.
Based on our proposed framework,  a distributed SMF zonotopic algorithm is developed.
Finally, simulation results indicate that our proposed algorithms are feasible to generate  estimation with relatively low conservativeness, and  effective facing linear time varying systems.

For future work, we aim to focus on extending our framework to non-linear systems, and consider the stability of our proposed framework.

\bibliographystyle{IEEEtran}

\bibliography{SetMembership_Filtering_Based_Cooperative_State_Estimation_for_MultiAgent_Systems}

\end{document}